\documentclass[10pt,twoside,BCOR7mm,DIV15,headinclude,footexclude,cleardoubleempty,idxtotoc]{scrartcl}

\usepackage[english]{babel}
\usepackage{graphicx}
\usepackage{hyperref}
\usepackage{scrpage2}
\usepackage{hyperref}
\usepackage{ifthen}

\makeatletter
\renewcommand{\@biblabel}[1]{}
\renewcommand{\@cite}[2]{%
{#1\ifthenelse{\boolean{@tempswa}}{,#2}{}}}
\makeatother

\hypersetup{breaklinks=true
,colorlinks=true,linkcolor=black,urlcolor=blue
,citecolor=black}

\ofoot{\thepage}
\ifoot{}

\setheadsepline{1pt}

\setkomafont{pagehead}{\normalfont\sffamily}
\setkomafont{pagenumber}{\normalfont\rmfamily}

\usepackage{booktabs}
\usepackage{amsmath}
\usepackage{amssymb}
\usepackage{multicol}
\usepackage{float}

\makeatletter
\newcommand{\listofcontributions}{\@starttoc{con}}

\newcommand{\l@contribution} {\@dottedtocline{1}{1.5em}{2.3em}}
\makeatother

\newenvironment{contribution}{
\setcounter{section}{0}
\setcounter{figure}{0}
\setcounter{table}{0}
\begin{flushleft}
{\em Clumping in Hot Star Winds \\
W.-R.\ Hamann, A.\ Feldmeier \& L.\ Oskinova, eds.\\
Potsdam: Univ.-Verl., 2007 \\
URN: http://nbn-resolving.de/urn:nbn:de:kobv:517-opus-13981
} 
\end{flushleft}
}{
\newpage
\lehead{}
\rohead{}
}

\begin{document}

\setlength{\baselineskip}{2.5ex}

\begin{contribution}

\lehead{R.\ Hirschi}

\rohead{Impact of reduced mass loss rates}

\begin{center}
{\LARGE \bf The impact of reduced mass loss rates on the evolution of
massive stars}\\
\medskip

{\it\bf Raphael HIRSCHI}\\

{\it University of Keele, United Kingdom}

\begin{abstract}
Mass loss is a very important aspect of the life of massive stars. After
briefly reviewing its importance, we discuss the impact of the recently
proposed downward revision of mass loss rates due to clumping (difficulty to form
Wolf-Rayet stars and production of critically rotating stars). 
Although a small reduction might be allowed, large reduction factors around ten are disfavoured. 

We then discuss the possibility of significant mass loss at very
low metallicity due to stars reaching break-up velocities and especially
due to the metal enrichment of the surface of the star via rotational
and convective mixing. This significant mass loss may help the first very 
massive stars
avoid the fate of pair-creation supernova, the chemical signature of
which is not observed in extremely metal poor stars. The chemical
composition of the very low metallicity winds is very similar to that of
the most metal poor star known to date, HE1327-2326 and offer an
interesting explanation for the origin of the metals in this star. 

We also discuss the importance of mass loss in the context of long and
soft gamma-ray bursts and pair-creation supernovae.
Finally, we would like to stress that mass loss in cooler parts
of the HR-diagram (luminous blue variable and yellow and red supergiant stages) 
are much more uncertain than in the hot part. More work needs to be done in 
these areas to better constrain the
evolution of the most massive stars.

\end{abstract}
\end{center}

\begin{multicols}{2}

\section{Introduction}

Mass loss has a crucial impact on the evolution of massive stars. It
affects evolutionary tracks, lifetimes and surface abundances. It also
determines the population of massive stars (number of stars in each
Wolf-Rayet subtype for example). It influences the type of supernova at
the death of the star (SNII, Ib, Ic, or a pair-creation supernova) 
and the final remnant 
(neutron star or black hole). Mass loss releases 
matter and energy back into the interstellar medium in amounts comparable
to supernovae (for stars above 30 $M_\odot$). Finally, it affects the hardness of
the ionizing radiation coming from massive stars. It is therefore very
important to understand mass loss in order to understand and model the
evolution of massive stars.

\section{Impact of reduced mass loss rates at solar metallicity}
The concept of clumping is not new (see contribution from Moffat in this volume, or 
a general review from Kudritzki \& Puls \cite{KP00ARAA}). However, new observations
suggest clumping factors leading to downward revision of mass loss rates between three
to ten or even more for massive stars (Bouret et al.\ \cite{BLH05}, Fullerton et al.\ \cite{FMP06}). 
Here we discuss the implications of
a reduction factor around ten. A 120 $M_\odot$ star, using current mass loss prescriptions at solar
metallicity 
(Vink et al.\ \cite{VKL00,VKL01}, Kudritzki \& Puls \cite{KP00ARAA}), 
loses on average $2\cdot 10^{-5}\,M_\odot$ per year. The lifetime of a 120 $M_\odot$ star is about
2.5 million years. This implies that, on the main sequence, a 120 $M_\odot$ star loses
approximately 50 $M_\odot$. If mass loss rates are reduced by a factor 10, such a star would
only lose 5 $M_\odot$. The question is then, how to produce a WR star with such low mass loss
rates? The first possibility is that mass loss is high in other evolutionary stages,
like the luminous blue variable (continuum-driven winds in LBVs, Smith \& Owocki \cite{SO06}) 
and the red supergiant (RSG) stages. Mass loss
rates are harder to determine in these two stages and therefore uncertainties are
still large. Nevertheless clumping may also affect mass loss determination in other
stages. Another possibility would be that all massive stars are in close binary
systems (Kobulnicky et al.\ \cite{KFK06}). However, if this were true, then it would be hard to produce the
many RSG stars observed. Furthermore, the fraction of Wolf-Rayet stars in close
binary systems in the Magellanic Clouds is found to be only 30-40\% 
(Foellmi et al.\ \cite{FMG03a,FMG03b}). 
This means that single stars must still be able to lose enough mass to become
WRs on their own. The last possibility discussed here is that rotation (possibly
coupled to magnetic fields) induces such a strong mixing that WRs are produced by
mixing rather than mass loss (Maeder \cite{Ma87}, Yoon \& Langer \cite{YL05}). This scenario works only
for fast rotators, which represent only a small fraction at $Z_\odot$ 
and therefore this cannot produce all the WR stars observed. 

Another important impact of strongly reduced mass loss rates is that it implies
that the angular momentum loss is much weaker. This would lead to many critically
rotating stars near the end of the main sequence, similar to Be stars. This in turn would
lead to an increase in mass loss rates, which could possibly
compensate for a modest reduction factor. Additional models are necessary to give a quantitative
answer. 
More interestingly, mass loss would become strongly anisotropic (Maeder \& Desjacques \cite{MD01}) 
and possibly produce disks when the $\Omega$-limit is reached as is observed
around Be type stars (See e.g.\ the review by Porter \& Rivinius \cite{PR03}).
The lack of observations of
critically rotating very massive stars is an argument against mass loss being extremely low on the main 
sequence and high only later on during the LBV and RSG stages.
Note that rotation 
is able to compensate for a reduction factor 2 for mass loss since comparable number of WR stars 
are produced with enhanced mass loss rates (Meynet et al.\ \cite{MM94}) and with normal mass loss rates + rotation 
(Meynet \& Maeder \cite{ROTXI}). 

\section{Metallicity dependence}
The metallicity ($Z$) dependence of mass loss rates is usually included using the 
formula: 
$\dot{M}(Z)=\dot{M}(Z_\odot)(Z/Z_\odot)^{\alpha}$.
The exponent $\alpha$ varies between 0.5-0.6 
(Kudritzki \& Puls \cite{KP00ARAA}, Kudritzki \cite{Ku02}) and
0.7-0.86 (Vink and collaborators \cite{VKL01, VdK05}) for O-type and WR stars
(See Mokiem et al.\ \cite{MKV07} for a recent comparison between mass loss
prescriptions and observed mass loss rates). Until very recently, 
most models use at best the total metal content present at the surface of the star 
to determine the mass loss rate. However, the surface chemical composition becomes
very different from the solar mixture, due either to mass loss in the WR stage or by
internal mixing (convection and rotation) after the main sequence. It is
therefore important to know the contribution from each chemical species to opacity and
mass loss. Recent studies (Vink et al.\ \cite{VKL00,VdK05}) show that iron is the dominant
element concerning radiation line-driven mass loss for O-type and WR stars. 
In the case of WR stars, there is
however a plateau at low metallicity due to the contributions from light elements
like carbon, nitrogen and oxygen (CNO). 
In the RSG stage, rates generally used are still those of
Nieuwenhuijzen \& de Jager (\cite{NdJ90}). 
Observations indicate that
there is a very weak dependence of dust-driven mass loss on metallicity and 
that CNO elements and especially nucleation seed components like silicon and titanium are
dominant (Van Loon \cite{VL00}, \cite{VL06}, Ferrarotti \& Gail \cite{FG06}). 
See van Loon et al.\ (\cite{VL05}) for recent mass loss rate prescriptions in the 
RSG stage. 
In particular, the ratio of carbon to oxygen is important to determine which
kind of molecules and dusts form. 
If the ratio of carbon to oxygen is larger than one, then carbon-rich dust would
form, and more likely drive a wind since they are more opaque than oxygen-rich
dust at low metallicity (H\"ofner \& Andersen \cite{HA07}). 

In between the hot and cool parts of the HR-diagram, mass
loss is not well understood. Observations of the LBV stage indicate that several solar
masses per year may be lost and there is no indication of a metallicity dependence.
Chromospheric activity could also play a role in stars having surface temperature
similar to the Sun. 
Thermally driven winds and pulsations are still other ways to lose mass.
Even though there are still large uncertainties in the dependence of 
the mass loss rates on metallicity in the cooler part of the HR-diagram, 
it is very useful to use models and observations at various metallicities. Indeed, 
clumping appears to be metallicity independent and therefore comparisons between
models and observations should yield the same conclusions at different metallicities.
Furthermore, using models at lower metallicity already give a very good estimate of what
the impact of clumping may be on the evolution of the star. The mass loss rate
is a factor 1.6-2.2 (depending on alpha) and 2.2-4.0 lower 
at the metallicity of the large and small Magellanic Clouds (LMC and SMC) 
respectively. Comparing models calculated at the SMC metallicity with observations
at solar metallicity shows the impact of a reduction factor around three. 

Several groups recently computed massive star models and compared them to observed populations
around solar metallicities. Here we present a few of them. 
Meynet \& Maeder (\cite{ROTXI}) compare the ratio or WR to O-type stars using 
$\alpha$=0.5 for O-type star and no metallicity
dependence for WR stars. They find that rotating models better reproduce the WR/O
ratio and also the ratio of type Ib+Ic to type II supernova as a function of metallicity compared
to non-rotating models, which underestimate these ratios. Reducing the mass loss rates
by even a factor two would not fit the observations as well as with the current mass
loss prescriptions. Eldridge \& Vink (\cite{EV06}) use mass loss rates dependent on metallicity in the WR
stage and find a better agreement with observations for the WC/WN ratio compared to
metallicity independent mass loss rates in the WR stage. Again, reducing the mass loss
rates by a factor 2 or more would not fit the data better than with the current mass
loss prescriptions. However Vanbeveren et al.\ (\cite{VVB07}) includes binary stars in the
comparison and find a good fit with a mass loss rate reduced by a factor two. 

Including all the arguments discussed above, from the current stellar evolution
point of view the observations of the populations of massive stars
would not be better reproduced with
mass loss rate prescriptions reduced by a factor greater than two.

\section{First stellar generations}
As we saw in the previous sections, mass loss plays a crucial in the evolution of
solar metallicity stars. In this section, we discuss the importance of mass
loss on the evolution of the first stellar generations. The first massive stars died
a long time ago and will probably never be detected directly 
(see however Scannapieco et al.\ \cite{SMWH05}).
There are nevertheless indirect observational constraints on the first stars coming
from observations of the most metal poor halo stars (Beers \& Christlieb \cite{BC05}). The
first stars are very important because they took part in the re-ionisation of the
universe at the end of the dark ages (roughly 400 million years after the Big Bang).
They are therefore tightly linked to the formation of the first structures in the
universe and can provide valuable information about the early evolution of the
universe.
The first stellar
generations are different from solar metallicity ($Z_\odot$) stars due to their low metal content
or absence of it. First, very low-$Z$ stars are more compact due to lower opacity.
Second, {\it metal free} stars burn hydrogen in a core, which is denser
and hotter. This implies that the transition between core hydrogen and helium burning is much
shorter and smoother. Furthermore, hydrogen burns via
the pp-chain in shell burning. {\it These differences make the metal free (first) stars different from the
second or later generation stars!} (Ekstr\"om \& Meynet \cite{EM07}). Third, mass loss is
metallicity dependent (at least for radiation-driven winds) and therefore mass loss is expected to become very small at very
low metallicity. Finally, the initial mass function of the first stellar generations is
expected to be top heavy below a certain threshold 
(Bromm \& Loeb \cite{BL03}). 

Mass loss is expected to be very small. What could change this expectation? An
additional mechanism or the chemical enrichment of the envelope of the star are two
possible ways to increase mass loss at very low $Z$. Models of metal free stars
including the effect of rotation (Ekstr\"om et al.\ \cite{EMM05}) show that stars may lose up to
10 \% of their initial mass due to the star rotating at its critical limit (also
called break-up limit). 
The mass loss due to the star reaching the critical
limit is non-negligible but at the same time not important enough to change
drastically the fate of the first generation stars. 

The situation is very different at very low but non-zero metallicity 
(Meynet et al.\ \cite{MEM06} and Hirschi \cite{H07}). 
The total mass of an 85\,$M_\odot$ model at $Z=10^{-8}$ is shown in Fig.
\ref{hirschi:kip85} with the top solid line. This model, like metal free models,
loses around 5\% of its initial mass when its surface reaches break-up velocities in
the second part of the main sequence. At the end of core H-burning,
the core contracts and the envelope expands, thus decreasing the surface
velocity and its ratio to the critical velocity. The mass loss rate becomes
very low again until the star crosses the HR diagram and reaches the RSG
stage. At this point the convective envelope dredges up CNO elements to
the surface increasing its overall metallicity. The total metallicity, $Z$, is 
used in this model (including CNO elements)
for the metallicity dependence of the mass loss.
\begin{figure}[H]
\begin{center}
\includegraphics[width=\columnwidth]{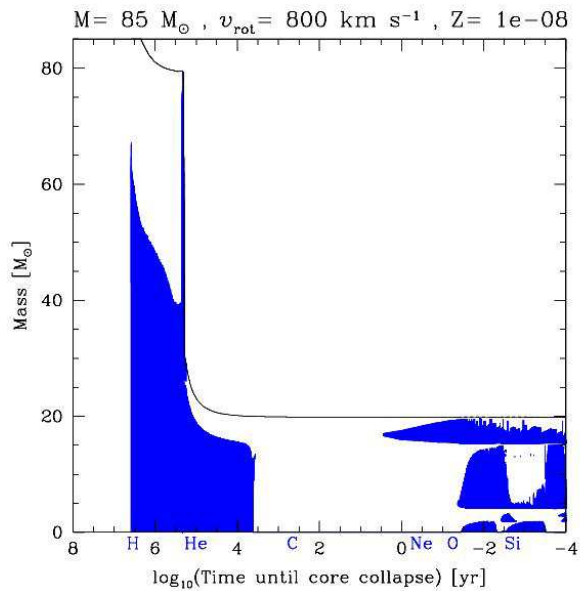}
\caption{Structure evolution diagram of the 85 $M_\odot$ model at $Z=10^{-8}$.
Coloured areas corresponds to convective zones along the lagrangian mass coordinate as a
function of time left until the core collapse. The top solid line shows the total mass of the
star. The burning stage abbreviations are given below the time axis.
\label{hirschi:kip85}}
\end{center}
\end{figure}
\noindent Therefore depending on how much CNO is brought up to the surface, the
mass loss becomes very large again. The CNO brought to the surface
comes from primary C and O produced in He-burning. Rotational and convective 
mixing brings these elements
into the H-burning shell. A large fraction of the C and O is then 
transformed into primary nitrogen via the CNO cycle. 
Additional convective and rotational
mixing is necessary to bring the primary CNO to the surface of the star.
The whole process is complex and depends on mixing. Multi-dimensional models
would be very helpful to constrain mixing between the hydrogen and carbon rich
layers, which releases a large amount of energy and strongly affects the structure of
the star.

The strongest mass loss occurs in these models in the cooler part of the HR
diagram. Dust-driven winds appear to be metallicity independent as long as C-rich dust
can form. For this to occur, the surface effective temperature needs to be low enough
(log(T$_{\rm eff})<3.6$) and carbon needs to be more abundant than oxygen. Note that
nucleation seeds (probably involving titanium) are still necessary to form C-rich dust. 
It is not clear
whether extremely low-$Z$ stars will reach such low effective temperatures. This
depends on the opacity and the opacity tables used in these calculations did not account for the
non-standard mixture of metals (high CNO and low iron abundance). It is interesting to note that the wind of the 85 $M_\odot$ model
is richer in carbon than oxygen, thus allowing C-rich dust to form if nucleation seeds are
present. There may also be
other important types of wind, like chromospheric activity-driven, pulsation-driven, 
thermally-driven or continuum-driven winds.

\begin{figure}[H]
\begin{center}
\includegraphics[width=\columnwidth]{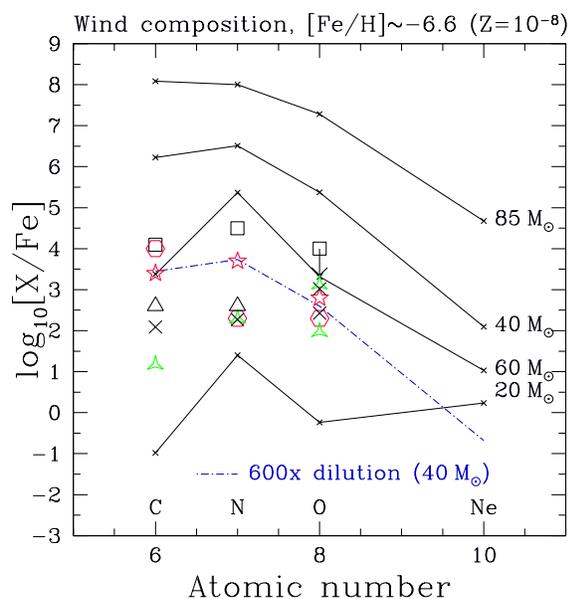}
\caption{Composition in [X/Fe] of the stellar wind for the $Z=10^{-8}$
models (solid lines).
For HE1327-2326 ({\it red stars}), the best fit for the
CNO elements is
obtained by diluting the composition of the wind of the 40 $M_\odot$
model by a factor 600 (see Hirschi \cite{H07} for more details).\label{hirschi:cempw}}
\end{center}
\end{figure}
Significant mass loss in very low-$Z$ massive stars offers an interesting
explanation for the strong enrichment in CNO elements of the most metal
poor stars observed in the halo of the galaxy 
(see Meynet et al.\ \cite{MEM06} and Hirschi \cite{H07}). 
The most metal poor stars known to date, 
HE1327-2326 (Frebel et al.\ \cite{Fr06}) is characterised by very high N, C and O abundances,
high Na, Mg and Al abundances, a weak s--process enrichment and depleted
lithium. The star is not evolved so has not had time to bring
self--produced CNO elements to its surface and is most likely a subgiant.
By using one or a few SNe and using a very large mass cut, 
Limongi et al.\ (\cite{LCB03}) and Iwamoto et al.\ (\cite{IUTNM05}) are
able to reproduce the abundance of most elements. 
However they are not
able to reproduce the nitrogen surface abundance of
HE1327-2326 without rotational mixing. 
A lot of the features of this star are similar to the properties of the
stellar winds of very metal poor rotating stars. 
HE1327-2326 could therefore have formed
from gas, which was mainly enriched by stellar winds of rotating very low
metallicity stars. In this scenario, a first generation of stars 
(PopIII) 
pollutes the interstellar medium to very low metallicities
([Fe/H]$\sim$-6). Then a PopII.5 star 
(Hirschi \cite{paris05}) like the 
40 $M_\odot$ model calculated here
pollutes (mainly through its wind) the interstellar medium out of
 which HE1327-2326 forms.
This would mean that HE1327-2326 is a third generation star.
In this scenario, 
the CNO abundances are well reproduced, in particular that of
nitrogen, which according to the new values for a subgiant from Frebel et al.\ (\cite{Fr06})
is 0.9 dex higher in [X/Fe] than oxygen. 
This is shown in Fig. \ref{hirschi:cempw} where the abundances of HE1327-2326 are
represented by the red stars and the best fit is 
obtained by diluting the composition of the wind of the 40 $M_\odot$
model by a factor 600. When the SN
contribution is added, the [X/Fe] ratio is usually lower for nitrogen
than for oxygen. 
Although the existence of a lower limit for the minimum metallicity $Z$ for low mass stars
to form is still under debate,
it is interesting to note that the very high CNO yields of the 
40 $M_\odot$ stars brings the total
metallicity $Z$ above the limit for low mass star formation
obtained in Bromm \& Loeb (\cite{BL03}).

\section{Gamma-ray bursts and pair-creation supernovae}
Long and soft gamma-ray bursts (GRBs) have now been firmly connected to the death
of type Ic supernovae (see Woosley \& Bloom \cite{WB06} for a recent review). In one of
the most promising models, the collapsar model (Woosley \cite{W93}), GRB progenitors
must form a black hole, lose their hydrogen rich envelope (become a WR) and retain enough 
angular momentum in their core during the pre-supernova stages. 
The strong mass loss discussed in the previous section make it possible
for single massive stars in the first stellar generations to become WR stars and
even to retain enough angular momentum to produce a GRB (Hirschi \cite{H07}). 
A wider grid of models at metallicities around solar shows that the
rate of fast rotating WO stars is compatible with the rate of GRBs with an upper limit
around the LMC metallicity, in agreement with observations (Hirschi et al.\ \cite{grb05}). 
More recent models
including the effects of magnetic fields (Yoon \& Langer \cite{YL05}) show that another mechanism is
possible to produce GRBs at low $Z$. This mechanism is the quasi-chemical
evolution of very fast rotating massive stars. In this scenario, WR stars are
produced by mixing and not mass loss. This last scenario however does not predict
GRB at metallicities equal or higher than the SMC. This upper limit is too low
compared to recent observations (Fruchter et al.\ \cite{Fru06}). 
Taking into account the anisotropy in the wind of these fast rotating stars 
(Meynet \& Maeder \cite{AM07})
may help reduce the discrepancy between models and observations. 
Note that the 
downward revision of solar metallicity (Asplund et al.\ \cite{AGS05}) 
may also help resolve the problem.

Apart from GRBs, pair-creation supernovae (PCSNe) are very energetic explosions,
which could be observed up to very high redshifts 
(Scannapieco et al.\ \cite{SMWH05}).
PCSNe are expected
to follow the death of stars in the mass range between 100 and 250 $M_\odot$, assuming
that they do not lose a significant fraction of their mass during the
pre-supernova stages. Amongst the very first stars formed in the Universe, one
expects to have PCSN due to the lack of mass loss and to the low opacity
unable to stop the accretion on the star during its formation.
However, the EMP stars observed in the halo of the galaxy do not show the
peculiar chemical signature of PCSN (strong odd-even effect, see Heger \& Woosley \cite{HW02}). 
This means that either too few or even no 
PCSN existed. One possible explanation to avoid the production of very
low-$Z$ or metal free PCSNe is the strong mass loss in the cool part of the HR
diagram due to the surface enrichment in CNO elements induced by rotational and 
convective mixing (see previous section) or the star reaching the 
$\Omega \Gamma$-limit (Ekstr\"om \& Meynet \cite{EM07}).

\bigskip{\em Acknowledgments.} I wish to thank G. Meynet and S. Ekstr\"om for their help during the 
preparation of this review.

%
%
%

\end{multicols}
\end{contribution}


\end{document}